\documentclass[conference]{IEEEtran}
\IEEEoverridecommandlockouts
\usepackage{cite}
\usepackage{amsmath,amssymb,amsfonts}
\usepackage{algorithmic}
\usepackage{graphicx}
\usepackage{textcomp}
\def\BibTeX{{\rm B\kern-.05em{\sc i\kern-.025em b}\kern-.08em
    T\kern-.1667em\lower.7ex\hbox{E}\kern-.125emX}}
\begin{document}

\title{Novel Coronavirus COVID-19 Strike on Arab Countries and Territories: A Situation Report I\\
}

\author{\IEEEauthorblockN{Omar Reyad}
\IEEEauthorblockA{\textit{College of Computing and Information Technology} \\
\textit{Shaqra University, Saudi Arabia}\\
\textit{Faculty of Science} \\
\textit{Sohag University, Egypt} \\
Email: ormak4@yahoo.com}
}

\maketitle

\begin{abstract}
The novel Coronavirus (COVID-19) is an infectious disease caused by a new virus called COVID-19 or 2019-nCoV that first identified in Wuhan, China. The disease causes respiratory illness (such as the flu) with other symptoms such as a cough, fever, and in more severe cases, difficulty breathing. This new Coronavirus seems to be very infectious and has spread quickly and globally. In this work, information about COVID-19 is provided and the situation in Arab countries and territories regarding the COVID-19 strike is presented. The next few weeks main expectations is also given.
\end{abstract}

\begin{IEEEkeywords}
Coronavirus, COVID-19, Arab Countries.
\end{IEEEkeywords}

\section{Introduction}  \label{intro}
\begin{figure*}
  \centering
\begin{center}
\includegraphics[width= 0.88\textwidth]{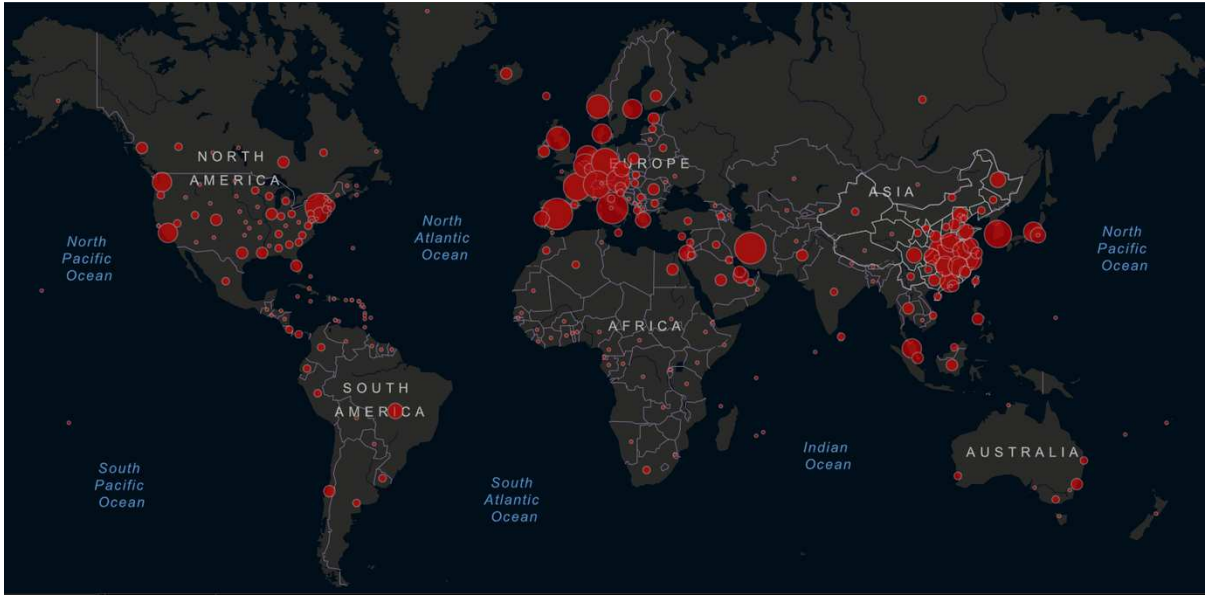}
\end{center}
\caption{Coronavirus COVID-19 Global Cases CSSE JHU 2020}
\label{Fig-1}       
\end{figure*}

The well known Coronaviruses such as MERS-CoV, SARS-CoV and COVID-19 are a group of viruses that infects both birds and mammals which meaning that they are transmitted between people and animals. These set of Coronaviruses cause infections that are related to the common cold and flu in humans where symptoms vary according to the infected species \cite{c1,c2}. The COVID-19 has reported being a novel Coronavirus of a typical pneumonia since the date 31/12/2019. The COVID-19 started in Wuhan city in China and then spread around the world very fast. Covid-19 is considered as the second Coronavirus outbreak that affects the Middle East region, following the MERS-CoV which was reported in Saudi Arabia in the year 2012. United Arab Emirates (UAE) was the first Middle East Arab country to report a Coronavirus-positive case, following the Wuhan city Coronavirus outbreak in China. Recently, on 11/03/2020, the World Health Organization (WHO) stated that the global COVID-19 outbreak is a pandemic because of the speed and scale of transmission of the virus. From the 195 countries in the world today, there are more than 266,100 Coronavirus total cases reported to Coronavirus resource center until now \cite{c3}. Moreover, the number of total deaths are more than 11,200 cases and the number of total recovered are more than 87,300 cases. Figure \ref{Fig-1} shows the Coronavirus COVID-19 global cases presented by the center for systems science and engineering (CSSE) at Johns Hopkins University (JHU) up-to-the-date 20/03/2020 \cite{c4}.

In this work, the up-to-date information about COVID-19 is provided and the situation in Arab countries and territories regarding the COVID-19 outbreak is presented.

The rest of the paper is organized as follows: In Section \ref{sec:1}, we presented recent related works on COVID-19. The COVID-19 basics are discussed in Section \ref{sec:2}. The situations in Arab countries and territories are discussed in Section \ref{sec:3} while conclusions are given in Section \ref{sec:4}.

\section{Related Works} \label{sec:1}
Several researchers were working under difficult and dangerous conditions to bring out new Coronavirus disease (COVID-19) content. In \cite{c5}, a modeling study on now-casting and forecasting the potential domestic and international spread of the 2019-nCoV outbreak originating in Wuhan, China is presented. In \cite{c6}, mass gathering events and reducing further global spread of COVID-19 is presented. An analysis of International Health Regulations annual report data from $182$ countries is given in \cite{c7}. Momentous decisions and many uncertainties of COVID-19 in Italy are proposed in \cite{c8}. In \cite{c9}, the resilience of the Spanish health system against the COVID-19 pandemic is discussed. The day level forecasting for Coronavirus disease (COVID-19) spread analysis is proposed in \cite{c10}. In \cite{c11}, authors introduced essential knowledge about COVID-19 and nosocomial infection in dental settings and provides recommended management protocols for dental practitioners and students in (potentially) affected areas.

\section{COVID-19 Basics} \label{sec:2}
The persons who are infected with Coronavirus COVID-19 may not know that they have symptoms of COVID-19 because these symptoms are similar to a cold or flu common symptoms. Symptoms of COVID-19 may take up to $14$ days to appear after the person exposure to COVID-19 virus. These days are the longest known infectious period for this disease believed by experts. 
 
COVID-19 symptoms may have included:

\begin{itemize} 
\item[$\ast$] \hspace {.2em} fever,
\item[$\ast$] \hspace {.2em} cough,
\item[$\ast$] \hspace {.2em} difficulty breathing,
\item[$\ast$] \hspace {.2em} pneumonia in both lungs.
\end{itemize}
    
There are some basic protective measures against the COVID-19 virus available on the WHO website and through your national and local public health authority. Most of the people who become infected experience moderate illness and heal, but it can be more severe for some cases and for other cases infection can lead to death. To protect others and take care, the following instructions is very helpful: 

\begin{enumerate}
\item \hspace {.2em} Regularly and thoroughly clean your hands with an alcohol-based hand rub or simply wash them with soap and water for 20 second.
\item \hspace {.2em} Maintain at least $1$ meter distance between a yourself and anyone who is coughing or sneezing beside you.
\item \hspace {.2em} Avoid touching the eyes, nose and mouth before cleaning it carefully.
\item \hspace {.2em} Make sure that you and the people around you are following good respiratory hygiene. Covering your mouth and nose with your bent elbow or tissue when you cough or sneeze. Then, get rid of the used tissue immediately.
\item \hspace {.2em} Stay at home if you feel diseased. If you have a fever, cough and difficulty breathing, seek medical attention and call in advance. 
\item \hspace {.2em} Follow the directions of your local health authority.
\end{enumerate}   

\section{The Situation in Arab Countries} \label{sec:3}
\begin{figure*}
  \centering
\begin{center}
\includegraphics[width= 0.88\textwidth]{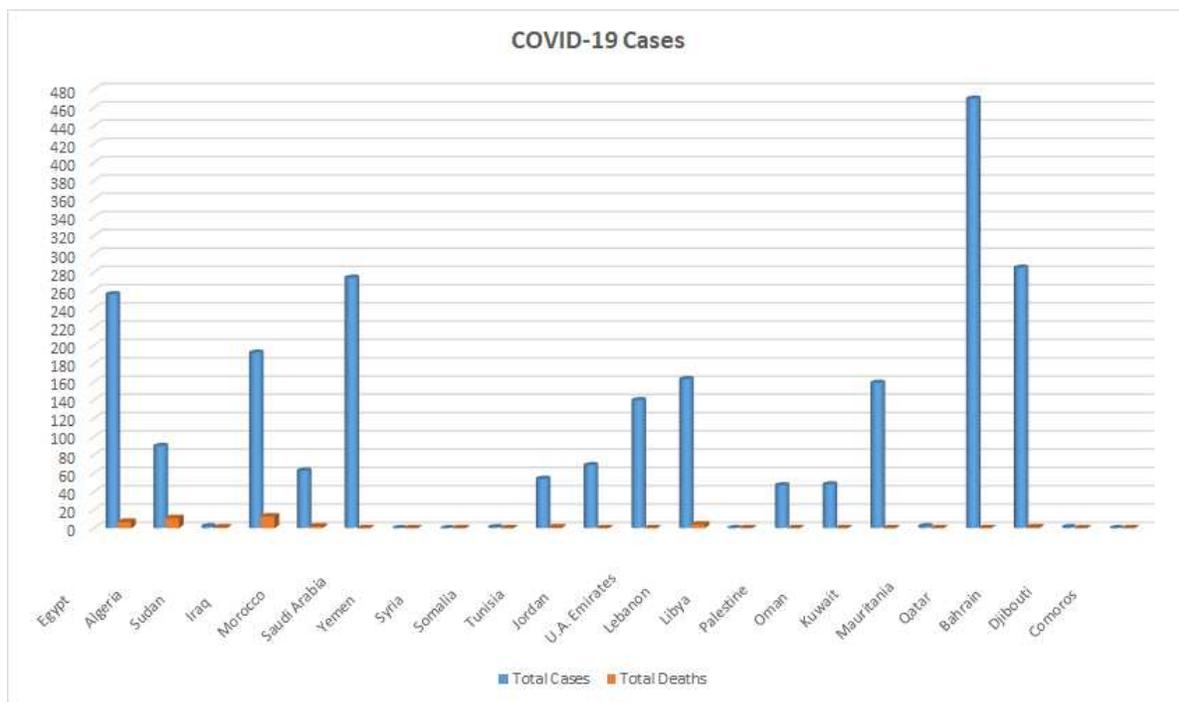}
\end{center}
\caption{COVID-19 Total Cases}
\label{Fig-2}       
\end{figure*}

The new COVID-19 virus has had a greater proliferation in a shorter amount of time than its predecessors SARS and MERS. However, the percentage of deaths remains much lower. Nearly, about $10\%$ of those infected died during the SARS outbreak, while $35\%$ died during MERS. Among the countries with Coronavirus in the Arab world, Qatar has the highest number of Coronavirus cases, followed by Bahrain and Saudi Arabia as shown in Table \ref{Stat} and Figure \ref{Fig-2}. Other Arab world countries that are currently free from Coronavirus include Syria and Comoros while Libya and Yemen have suspected cases.

As a precautionary measure, almost all Arab countries has temporarily banned entry to people and tourists from Coronavirus-affected countries and other countries as well. Most of Arab countries also close their borders and stop flights to/from other countries for different period of time. 

In the next few weeks there are more than one scenario that may happen in the Arab countries as listed here:

\begin{enumerate}
\item \hspace {.2em} The best and hopefully scenario that the Arab world countries will be able to succeed in dealing with the outbreak and preventing widespread infection among its population especially with the help of vaccine(in case it appears).
\item \hspace {.2em} The infection rate will continue in growth and will soon reach the peak and start return back with the cost of many infected cases and deaths. 
\item \hspace {.2em} The worst case that the infection take more time and many cases with COVID-19 will still appear with in the next few months. This is worst and nightmare for many Arab countries with limited resources.
\item \hspace {.2em} The COVID-19 vaccine will appear in the perfect time to overcome all this bad expectations.
\end{enumerate}

{\begin{table} [htbp]
\caption{Arab Countries and Territories Cases of COVID-19}
\label{Stat}
\begin{center}
\begin{tabular}{| l | c | c | } \hline
\textbf{Country}        & \textbf{Total Cases} & \textbf{Total Deaths} \\ \hline
Egypt                  & 256  & 7     \\ \hline
Algeria                & 90   & 11       \\ \hline
Sudan                  & 2    & 1     \\ \hline
Iraq                   & 192  & 13    \\ \hline
Morocco                & 63   & 2     \\ \hline
Saudi Arabia           & 274  & 0     \\ \hline
Yemen                  & NA   & NA     \\ \hline
Syria                  & NA   & NA      \\ \hline
Somalia                & 1    & 0     \\ \hline
Tunisia                & 54   & 1     \\ \hline
Jordan                 & 69   & 0     \\ \hline
U.A. Emirates          & 140  & 0     \\ \hline
Lebanon                & 163  & 4     \\ \hline
Libya                  & NA   & NA     \\ \hline
Palestine              & 47   & 0    \\ \hline
Oman                   & 48   & 0     \\ \hline
Kuwait                 & 159  & 0    \\ \hline
Mauritania             & 2    & 0    \\ \hline
Qatar                  & 470  & 0    \\ \hline
Bahrain                & 285  & 1    \\ \hline
Djibouti               & 1    & 0    \\ \hline
Comoros                & NA   & NA    \\ \hline
\end{tabular}
\end{center}
\end{table}}

\section{Conclusions and Future Work}  \label{sec:4}
The novel Coronavirus COVID-19 is declared as an international epidemic on March 2020. The disease causes respiratory illness (such as the flu) with other symptoms such as a cough, fever, and in more severe cases, difficulty breathing. This new Coronavirus seems to be very infectious and has spread quickly and globally. In this work, information about COVID-19 is provided and the situation in Arab countries and territories regarding the COVID-19 strike is presented. The next few weeks main expectation scenarios is also given.

\end{document}